\documentstyle[12pt,epsfig]{article} 

\begin{document} 
\title{Quantum Binary Decision for Driven Harmonic Oscillator} 
\author{Matteo G. A. Paris\\ 
Dipartimento di Fisica 'Alessandro Volta' \\
Universit\`a di Pavia and I.N.F.M. -- Unit\'a di Pavia \\ 
via A. Bassi 6, I-27100 Pavia, ITALY \\
{\tt MATTEO.PARIS@PV.INFN.IT}} 
\date{}
\maketitle
\abstract{
We address the problem of determining whether or not 
a harmonic oscillator has been perturbed by an external force. 
Quantum detection and estimation theory has been used in 
devising optimum measurement schemes. 
Detection probability has been evaluated for different initial 
state preparations of oscillator. The corresponding lower bounds 
on minimum detectable perturbation intensity has been evaluated and a 
general bound for random phase perturbation has been also induced. 
} 
\section{Introduction} 
The harmonic oscillator is a relatively simple model, which is 
widely utilized in many fields of physics. Indeed, it provides a 
satisfactory description of a large number of very different 
physical systems. This is true also in a quantum mechanical framework, 
where the harmonic oscillator plays a crucial role. Its spectrum of 
eigenvalues, in fact, is infinite, discrete and bounded from below, thus 
representing a paradigm for any bounded oscillating system. \par 
Interesting physical features often comes with perturbation to 
harmonic behaviour, which are to be revealed from measurement 
performed on the system. This is the case, as an example, of 
oscillating electronic circuits \cite{he1} or of some large mass 
viewed as a gravitational antenna \cite{web,hol,bon}. Also, a single mode 
radiation field is modeled on harmonic oscillator, and many optical 
devices act as driving terms in the dynamical evolution \cite{wol}. 
\par 
Quantum mechanically, different measurements 
lead to different informations about a physical system, as each 
observable shows up only an aspect of the state under examination 
\cite{ste,par,opt}. Therefore, it is a matter of interest to analyze 
the different measurement processes, in order to find an optimized 
measurement scheme, which is capable to reveal perturbations as weak 
as possible. This project is a matter of quantum detection and 
estimation theory \cite{heL,hoL}, which regards the very general 
problem of extracting information on a physical system from measurement 
or a set of measurements. 
\par
In the present paper we address the binary decision theory for driven 
quantum harmonic oscillator. Let us consider an oscillator which is free 
to follow harmonic evolution, and possibly subjected to an external driving 
signal. After a fixed time $\tau$ we perform a measurement on the system, 
in order to check whether or  not the oscillator has been perturbed. 
We are going to deal with two questions: first, which is the best 
measurement one can perform, in order to reveal perturbations 
as weak as possible with the minimum probability of error ? 
And second, which is the minimum detectable perturbation intensity, 
depending on the initial state preparation of the oscillator ? 
\par 
We do not concern to any specific measurement device and we do not 
discuss the feasibility of optimized measurement. Rather, we attempt 
to derive a ultimate quantum limit on the detectable intensity of a 
perturbation, which depends only on the initial quantum state of the 
oscillator. \par 
The paper will be organized as follows. In Section \ref{s:qm} we set the 
proper quan\-tum measu\-rement theory frame\-work and illu\-strate the 
Ney\-man-Pear\-son 
stra\-tegy for binary decision. In Section \ref{s:qs} we consider different 
initial preparation states for the harmonic oscillator and derive the 
corresponding minimum detectable perturbation intensity. 
Section \ref{s:ou} closes the paper with some concluding remarks. 
\section{Quantum Detection Theory} 
\label{s:qm} 
\subsection{Quantum measurements} 
\label{sb:qm} 
In a quantum mechanical framework any measurement apparatus is a 
device, at least a mathematical one, which turns each quantum 
state $\hat\rho$ into a probability density distribution \cite{oza} 
\begin{equation} 
dP[\hat\rho] : \hat\rho \longrightarrow dP[\hat\rho](x) \qquad x\in {\cal X} 
\;,\label{basic} 
\end{equation} 
where $\cal X$ is some measurable space, where the possible outcomes of the 
measurement lie. This can be the real Borel set or a subspace of 
it. The measurement map is provided by trace operation 
\begin{equation} 
dP[\hat\rho](x)=\hbox{Tr} \left\{ \hat\rho \; d\hat\mu (x) \right\} 
\label{trace}\;, 
\end{equation} 
which assures propagation of convex linear combinations from density 
operators toward probabilities. The above formula is the Born's 
statistical rule \cite{bor}. It contains the whole probabilistic 
structure of quantum mechanics \cite{brv}. 
The Born rule leads to a genuine probability density distribution if the 
operator $d\hat\mu (x)$ satisfies the axioms for a 
{\it probability operator measure} (POM) \cite{mla}, namely it is 
nonnegative 
\begin{equation} 
d\hat\mu (x) \geq 0 
\label{none}\;, 
\end{equation} 
and it provides a resolution of identity on the set of possible outcomes 
\begin{equation} 
\int_{\cal X} d\hat\mu (x) = \hat {\bf 1} 
\label{norm}\;. 
\end{equation} 
Eq. (\ref{norm}) guarantees the probability density in Eq. (\ref{trace}) 
to be normalized, whereas positiveness of $d\hat\mu (x)$ also 
assures it is selfadjoint. \par 
Spectral, orthogonal resolution $d\hat E(z)$ of a selfadjoint operator 
\begin{eqnarray} 
\hat Z = \int_{\cal Z} z \; d\hat E(z)&\qquad &d\hat E(z)=|z\rangle\langle 
z|\; dz \nonumber \\ 
\hat Z |z\rangle =z |z\rangle &\qquad & \langle z|z' \rangle = \delta_{\cal Z} 
(z-z') 
\label{reso}\;, 
\end{eqnarray} 
provides a {\it projection valued measure} 
(PVM) which belongs to the class of POM. However, this is not the most general 
example. Also nonorthogonal projectors or overcomplete sets provides POM, 
namely available measurement scheme \cite{hev,wal,rip}.\par 
It is worth noting that a POM is necessarily (Naimark Theorem) \cite{nai} a 
partial trace of a PVM coming from a selfadjoint operator defined on a 
larger Hilbert space. The latter can be thought as the whole Hilbert space 
describing both, the system under examinations and the measurement 
apparatus \cite{cos}. As we are not going to deal with physical 
implementation of measurement we can restrict our attention on the Hilbert 
space of the examined system only. Thus, any measurement is properly 
described by a POM. 
\subsection{Driven Harmonic Oscillator} 
\label{sb:dha} 
The quantum mechanical description of harmonic oscillator is based on 
annihilation $a$ and creation $a^{\dag}$ operators 
\begin{equation} 
a= \sqrt{\frac{m\omega}{2}} x + i \sqrt{\frac{1}{2m\omega}} p 
\label{ac}\;, 
\end{equation} 
which form the number 
operator $\hat N = a^{\dag}a$. Due to commutation relation 
$[a,a^{\dag}]=1$, multiple applications of $a^{\dag}$ to the 
vacuum state leads to the Fock basis $|n\rangle = (n !)^{-1/2} a^{\dag n} 
|0\rangle$ which span the whole Hilbert space representing the possible levels 
of excitation for the oscillator. Number states represents also 
the eigenstates of the number operator, whose spectrum coincides with the set 
of the natural numbers ${\cal N} = 0,1,...$. 
The eigenstates of annihilation operator $a|\alpha\rangle = \alpha 
|\alpha\rangle$ constitute the overcomplete set of coherent states, 
$\alpha$ being the complex amplitude of the harmonic oscillations. 
Coherent states can also be obtained from the vacuum by the action of 
displacement operator $\hat D(\alpha)=\exp\{\alpha a^{\dag}-\bar{\alpha} a\}$ 
\cite{gla} 
\begin{eqnarray} 
|\alpha \rangle &=& \hat D(\alpha) |0\rangle 
\nonumber\\ 
\hat D(\alpha) \hat D(\alpha ') &=&\hat D(\alpha +\alpha ') 
\exp\left\{ \hbox{Im} (\alpha\alpha ')\right\} 
\label{discoh}\;. 
\end{eqnarray} 
Let us consider the Hamiltonian (natural unit $\hbar=1$) 
\begin{equation} 
{\bf H} = -\frac{1}{2m}\frac{d^2}{dx^2} + \frac{1}{2}m\omega^2 x^2 
+ F(t) x \label{ham}\;. 
\end{equation} 
It describes a classical harmonic oscillator of mass $m$ and 
frequency $\omega$ subjected to a time dependent driving force. 
In terms of annihilation and creation operator 
the Hamiltonian could be written as  \cite{lui} 
\begin{equation} 
{\bf H}= \omega a^{\dag} a + \frac{F(t)}{\sqrt{2m\omega}} (a^{\dag} + a) 
\label{ham1}\;. 
\end{equation} 
Finally, we adopt interaction (Dirac) picture to obtain 
\begin{equation} 
{\bf H}_I=\frac{F(t)}{\sqrt{2m\omega}} (a^{\dag} + a) 
\label{ham2}\;. 
\end{equation} 
Let now consider the initial state of the oscillator to be $\hat\rho_0$. 
If no driving force is present the final state, after a fixed 
evolution time $\tau$, is still $\hat\rho_0$. Otherwise, we have 
\begin{equation} 
\hat\rho_1 = \hat U \;\hat\rho_0 \;\hat U^{\dag} 
\label{evo}\;, 
\end{equation} 
where the evolution operator $\hat U= \exp\left\{i{\bf H}_I \tau\right\}$ 
could be written as a displacement operator \cite{dis} 
\begin{equation} 
\hat U \equiv \hat D (z) = \exp\left\{ za^{\dag} -\bar{z} a \right\} 
\label{u-d}\;, 
\end{equation} 
where 
\begin{equation} 
z=\frac{i\gamma\tau}{\sqrt{2m\omega}}\;, \qquad\quad 
\gamma=\int_0^{\tau} dt \; e^{i\omega t}\;F(t) 
\label{gamamzed}\;. 
\end{equation} 
The quantity $z$ represents the complex amplitude of the driving signal, 
whereas $|z|^2$ denotes the energy intensity of the perturbation, expressed 
in unit of the oscillator quanta $\omega$. 
\subsection{Neyman-Pearson Strategy for Binary Decision} 
\label{sb:nps} 
\begin{figure}[ht]
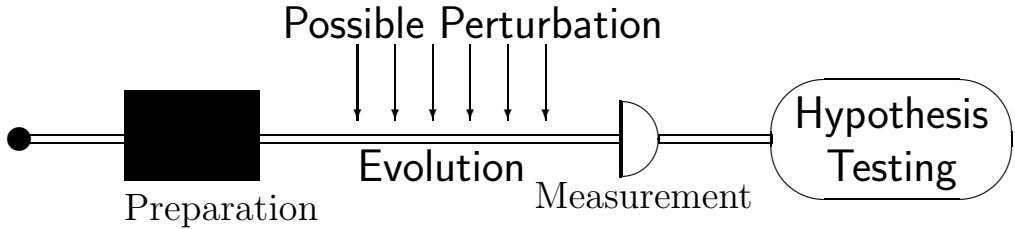
 
\input binfig1.pic
\caption{\sc 
outline of the detection scheme. The system is initially prepared in 
some fixed state and then it is left free to evolve. During the free evolution 
it could be subjected to an external driving signal. After a fixed time 
interval $\tau$ we perform some kind of measuremen on the system. From the 
outcome of such a measurement we have to infer which is the state of the 
system, in order to discriminate between the perturbation hypothesis and the 
null hypothesis.} 
\label{f:scheme} 
\end{figure}

Our goal is to determine whether or not the system has been perturbed. 
For this purpose we adopt a detection scheme as in Fig. \ref{f:scheme}. 
After the initial preparation the harmonic oscillator is left free to 
evolve for a fixed time $\tau$. Then, some kind of measurement 
$d\hat\mu (x)$ is performed. Starting from the outcomes of such a 
measurement we have to infer which is the state of the system, 
in order to discriminate between the following two hypothesis: 
\begin{itemize} 
\item[${\cal H}_0$:] No perturbation has been occurred during the time interval 
                    $\tau$, true if we infer $\hat\rho_0$; 
\item[${\cal H}_1$:] The system has been perturbed during the time interval 
                    $\tau$, true if we infer $\hat\rho_1$. 
\end{itemize} 
We denote by $P_{01}$ the probability of wrong inference, namely that 
one of inferring ${\cal H}_1$ when ${\cal H}_0$ is true. In hypothesis 
testing formulation this is usually referred to as {\it false alarm 
probability} \cite{leh}. Conversely, we denote by $P_{11}$ the {\it detection 
probability}, that is the probability of inferring ${\cal H}_1$ when 
it is actually true. \par
\par 
Now, which is the best measurement to discriminate between 
$\hat\rho_0$ and $\hat\rho_1$ ? \par 
If these two states are mutually orthogonal the problem has a trivial 
solution. It is a matter of measuring the observable for which 
$\hat\rho_0$ and $\hat\rho_1$ are eigenstates. However this is 
not our case, as displacing a state of the harmonic oscillator leads 
to a different kind of state. Only coherent states maintain their 
characteristic under displacement 
\begin{eqnarray} 
\hat D(z) |\alpha \rangle  &=& \hat D(z)\hat D(\alpha) | 0\rangle = 
\exp\left(\hbox{Im}[\alpha \bar{z}]\right) |\alpha +z \rangle 
\label{discoh1}\;. 
\end{eqnarray} 
However, coherent states constitute a nonorthogonal, 
overcomplete set by themselves. Thus, the above procedure cannot be 
applied in the present case, even for special initial states of the 
oscillator. \par 
In the following we consider nonorthogonal $\hat\rho_0$ and $\hat\rho_1$ and 
we focus our attention on oscillator initially prepared in a pure state 
$\hat\rho_0 = |\psi_0 \rangle\langle \psi_0 |$. As it can easily checked 
from Eq. (\ref{evo}) this means that also the perturbed state is a pure state 
$\hat\rho_1 = |\psi_1 \rangle\langle \psi_1 |$. \par 
The optimization problem can be analytically solved, for pure states, 
by adopting, the Neyman-Pearson criteria for binary decision \cite{nep}. 
The latter reads as follows. First, we have to fix a value for the 
false alarm probability $P_{01}$. Then, we have to find the measurement 
strategy $d\hat\mu (x)$ which maximizes the detection probability $P_{11}$. 
As a general definition, each measurement strategy which maximizes the 
detection probability $P_{11}$ for a fixed value of false alarm 
probability  $P_{01}$ 
is considered as a Neyman-Pearson optimized detection for binary 
hypothesis testing. It was shown by Helstrom \cite{heL} and Holevo 
\cite{hoL} that this very general 
problem could be reduced to solving the eigenvalue problem 
for the operator 
\begin{equation} 
d\hat\mu (x|\lambda ) = \hat\rho_1 -\lambda\hat\rho_0 
\label{opt}\;, 
\end{equation} 
which represents the optimized measurement scheme. In general it is a POM 
rather than a PVM. Nevertheless when, as it is here the case, the two signals
are linearly independent it has been proved by Kennedy \cite{ken,osa} that the 
optimum detection is described by a PVM.
The parameter $\lambda$ is a Lagrange multiplier. Different 
values of $\lambda$ correspond to different values of the false alarm 
probability, namely to a different Neyman-Pearson strategies. \par 
Once the eigenvalues problem for $d\hat\mu (x|\lambda )$ has been solved it 
results that only positive eigenvectors contribute to the detection 
probability $P_{11}$ \cite{heL,var,lax}. Thus the decision strategy is transparent: 
after a measurement of the quantity $d\hat\mu (x|\lambda )$ if the outcome is 
positive we infer perturbation hypothesis ${\cal H}_1$ is true. Conversely, 
we infer null hypothesis ${\cal H}_0$ when obtaining negative outcome. 
By expanding the eigenstates of $d\hat\mu (x|\lambda )$ in terms of 
$|\psi_0\rangle$ and $|\psi_1\rangle$ Lagrange multiplier $\lambda$ can be 
eliminated from the expression of detection probability which results 
\begin{equation} 
P_{11} = \left\{ 
\begin{array}{cr} 
\left[ \sqrt{P_{01} \kappa} + \sqrt{(1-P_{01})(1-\kappa)}\right]^2 
& 0\leq P_{01} \leq \kappa \\ 
 & \\ 
 1 & \kappa \leq P_{01} \leq 1 
\end{array} 
\right. 
\label{dprob}\;. 
\end{equation} 
In Eq. (\ref{dprob}) $\kappa$ denotes the square modulus of the overlap 
between perturbed and unperturbed state of the harmonic oscillator, 
in formula $\kappa = |{\cal O}[\psi_0,z]|^2$, where 
\begin{equation} 
{\cal O}[\psi_0,z] = \langle \psi_0 | \psi_1\rangle = 
\langle \psi_0 |\hat D(z) |\psi_0\rangle 
\label{overlap}\;. 
\end{equation} 
The overlap depends both on the initial state and on the perturbation 
amplitude. In the next Section we evaluate the quantity in 
Eq. (\ref{overlap}) for relevant kinds of initial state. \par 
\begin{figure}[htb]  
\psfig{file=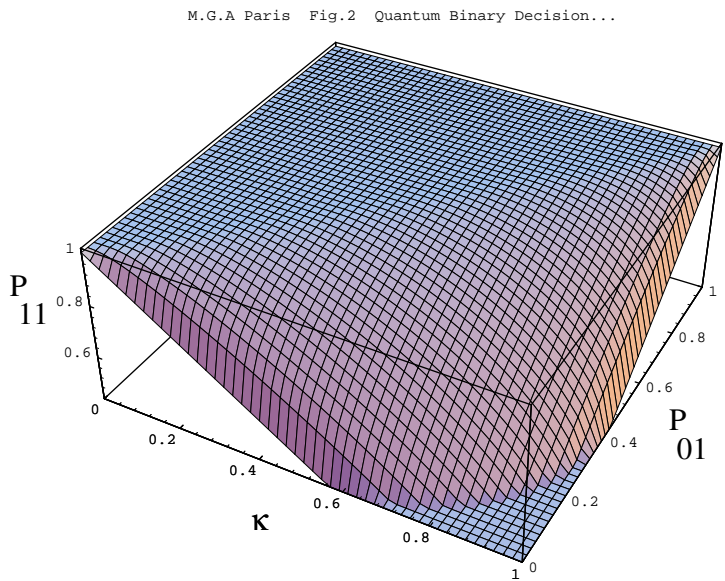,width=12cm}
\caption{\sc Three dimensional plot of the detection probability 
$P_{11}$ for Neyman-Pearson optimized detections as a function 
of the false alarm probability $P_{01}$ and the overlap $\kappa$ 
between the initial state and the perturbed one.} 
\label{f:basic} 
\end{figure} 
\par
In Fig. \ref{f:basic} we report the detection probability of 
optimized detection strategies as a function of the false alarm 
probability $P_{01}$ and the overlap strength parameter $\kappa$. It is 
obvious that if the overlap is small, it is easy to discriminate between 
the two states. Thus, it is possible to obtain strategies with large 
detection probability without paying the price of an also large false 
alarm probability. On the contrary, if the overlap becomes appreciable it 
is difficult to discriminate the states. In the limit of complete overlap 
the perturbed and the unperturbed states become indistinguishable. 
Detection probability is now equal to false alarm probability and 
the decision strategy is just a matter of guessing after each random 
measurement outcome. 
\par 
Choosing a value for the false alarm probability is a matter of 
convenience, depending on the specific problem this approach would be 
applied. The maximum tolerable value for $P_{01}$ increases with the 
expected number of measurement outcomes, and conversely a very low rate 
detection scheme needs a very small false alarm probability. 
\section{Quantum Binary Decision for Harmonic Oscillator} 
\label{s:qs} 
Once an acceptable value of false alarm probability has been fixed 
and the oscillator has been prepared in some initial state 
$|\psi_0 \rangle$, the detection probability $P_{11}$ depends only 
on the perturbation intensity. A wise inference could be performed 
only when $P_{11} (z) \geq 1/2$, as only in this case the record 
$X=\{x_0,x_1, ..., x_N\}$ of experimental data contain usable information. 
Thus, the threshold value $P_{11} (z) = 1/2$ defines the minimum detectable 
perturbation for the so-prepared oscillator plus detector system. We will 
consider the intensity of the minimum detectable perturbation as the relevant 
parameter and we denote it by $|z_{min}|^2 = {\cal M}$. In the following 
Subsections we study the behaviour of $\kappa$ and ${\cal M}$ for different 
initial preparation $|\psi_0 \rangle$. 
\subsection{Coherent States} 
\label{sb:cs} 
For the oscillator prepared in a coherent state $|\alpha\rangle$ the 
overlap is given by 
\begin{equation} 
{\cal O}[\alpha,z] = \langle\alpha |\hat D (z)|\alpha\rangle = 
\exp\{ -\frac{1}{2}|z|^2\} \exp\{z\bar{\alpha}-\alpha\bar{z}\} 
\label{ovcoh}\;, 
\end{equation} 
and thus the overlap strength does not depend on the amplitude 
of the prepared coherent state 
\begin{equation} 
\kappa = \exp\{ -|z|^2\} 
\label{stcoh}\;. 
\end{equation} 
For zero false alarm probability the detection probability is given by 
$P_{11}(z)=1-\exp\{ -|z|^2\}$. When a small false alarm probability is set 
($P_{01} \leq \kappa$) the minimum detectable perturbation intensity 
is obtained by the inversion of the formula 
\begin{equation} 
\frac{1}{2} = \left[ \sqrt{P_{01} \kappa} + \sqrt{(1-P_{01}) 
(1-\kappa)}\right]^2 
\label{mincoh1}\;, 
\end{equation} 
that is, 
\begin{equation} 
{\cal M}=\log\left( \frac{2}{1+\sqrt{P_{01}(1-P_{01})}}\right) 
\label{mincoh2}\;. 
\end{equation} 
The minimum detectable intensity is independent on the initial 
coherent amplitude. Thus coherent states provide a stable oscillating 
system, however also difficult to monitor in its fluctuations. 
\subsection{Squeezed States} 
\label{sb:ss} 
Uncertainty principle set a lower bound for the product of fluctuations for 
two conjugated quantity. This implies a degree of freedom, namely that one 
can arbitrarily reduce the fluctuations in some variable upon increasing 
the fluctuations in the conjugated one. Indeed, squeezed states of the 
harmonic oscillator have been introduced as minimum uncertainty 
state for amplitude quadrature operators $\hat x_{\varphi} \propto 
a^{\dag}e^{i\varphi}+ a e^{-i\varphi} $ with phase dependent fluctuations. 
\par 
Squeezed state can be obtained by the coherent displacement of squeezed vacuum 
$|\alpha,\zeta\rangle = \hat D(\alpha) |0,\zeta\rangle$ \cite{sto,yue}. The latter 
is obtained from the vacuum by the action of squeezing operator 
$|0,\zeta\rangle=S(\zeta ) |0\rangle$, where 
\begin{equation} 
S(\zeta )= \exp\left\{ \frac{1}{2}\left[\zeta a^{\dag 2} - 
\bar{\zeta}a^2\right]\right\} 
\label{opsq}\;, 
\end{equation} 
and $\zeta = r \exp\{i2\psi\}$, with $r$ real. 
Squeezing a state implies the introduction of some energy. The mean 
excitation number of a squeezed vacuum is given by 
$\bar{n}_{sq} \equiv \langle\zeta ,0| 
\hat N |0,\zeta\rangle = \sinh^2 r$. \par 
As we have seen just above coherent amplitude does not cause any 
effect when subjected to displacement action. Therefore, we restrict 
our attention to squeezed vacuum $|0,\zeta\rangle$ which shows all the 
interesting phase dependent features related to squeezing. We also 
consider, for simplicity, a squeezed vacuum with 
squeezing phase equal to zero $\psi =0$. \par
\begin{figure}[t]  
\psfig{file=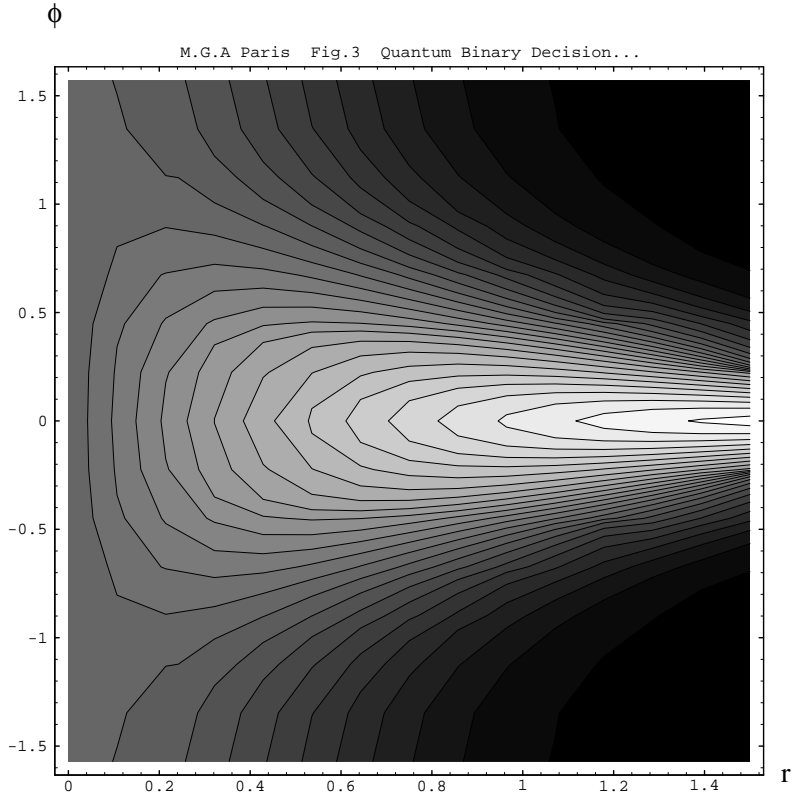,width=12cm}
\caption{\sc Contour plot of the overlap for the oscillator 
initially prepared in a squeezed vacuum. A perturbation with unit 
intensity is considered and the overlap is reported as a function of the 
squeezing parameter $r$ and the phase $\varphi$ of the perturbation.} 
\label{f:osq} 
\end{figure}

 
The overlap is given by 
\begin{equation} 
{\cal O}[r,z] = \exp\left\{-\frac{1}{2}|z|^2 \left[ 
\cosh 2r - \sinh 2r \cos^2\varphi 
\right]\right\} 
\label{ovsq}\;, 
\end{equation} 
where $\varphi\equiv\arg (z)$ is the phase of the perturbation. 
In Fig. \ref{f:osq} we report the overlap (\ref{ovsq}) for a unit intensity 
perturbation $|z|^2=1$ as a function of the squeezing parameter $r$ and the 
perturbation phase $\varphi$. 
In the two limiting cases $\varphi=0, \pi/2$ we have, for the overlap 
strength 
\begin{equation} 
\kappa_{0} = 
\exp\left\{ -2 |z|^2 \left[\bar{n}_{sq} +\frac{1}{2} - 
\sqrt{(\bar{n}_{sq} +1)\bar{n}_{sq}}\right] \right\} 
\label{k1sq0} 
\end{equation} 
\begin{equation} 
\kappa_{\pi/2} = 
\exp\left\{ -|z|^2 \left[ 2\bar{n}_{sq} +1 \right]\right\} 
\label{k1sqpi}\;. 
\end{equation} 
However, the perturbation phase is reasonably random, or unknown. Thus, a 
relevant parameter to be considered is also the phase averaged overlap 
strength  which is defined by 
\begin{equation} 
\bar{\kappa}  = \left| \int_{\pi}^{\pi} \frac{d\varphi}{2\pi} \; {\cal 
O}[r,|z|e^{i\varphi}] \right|^2 
\label{k2sq}\;. 
\end{equation} 
Inserting Eq. (\ref{ovsq}) in Eq. (\ref{k2sq}) leads to 
\begin{eqnarray} 
\bar{\kappa}&=& \exp\left\{ - |z|^2 \left[2\bar{n}_{sq} + 1 - 
\sqrt{(\bar{n}_{sq} +1)\bar{n}_{sq}}\right] \right\} 
\times 
\nonumber \\ 
&\times& 
\left[I_0 \left(\frac{1}{2}|z|^2  \sqrt{(\bar{n}_{sq} 
+1)\bar{n}_{sq}}\right)\right]^2 
\label{k3sq}\;, 
\end{eqnarray}
\par
\par 
with $I_0 (x)$ denoting a modified Bessel function of 
the first kind \cite{grd}. \par 
The minimum perturbation equation (\ref{mincoh1}) can be 
ana\-lytical\-ly sol\-ved for 
a fixed value of perturbation phase. We obtain 
\begin{equation} 
{\cal M}_{0} = 
\log\left( \frac{2}{1+\sqrt{P_{01}(1-P_{01})}}\right) 
\exp\{2r\} 
\label{minsq1}\;, 
\end{equation} 
where $e^{2r}=2\bar{n}_{sq}+1+2\sqrt{(\bar{n}_{sq} 
+1)\bar{n}_{sq}}$, and 
\begin{equation} 
{\cal M}_{\pi/2} = 
\log\left( \frac{2}{1+\sqrt{P_{01}(1-P_{01})}}\right) 
\frac{1}{2\bar{n}_{sq}+1} 
\label{minsq2}\;. 
\end{equation} 
From Eqs. (\ref{minsq1}) and (\ref{minsq2}) is apparent the strong 
effect of phase matching. When both phases, the perturbation one 
and the squeezing one, have the same value the overlap is strongly 
enhanced and thus the minimum detectable intensity increase 
(roughly linearly) with the 
increasing of the squeezing energy. On the contrary when the two phases 
are 
maximally mismatched, the overlap decreases with increasing energy of the 
initial states. Thus, the system becomes more and more sensitive to 
perturbation and minimum detectable intensity shows an inverse scaling 
with the initial preparation energy. \par 
\begin{figure}[htb]  
\begin{center}\psfig{file=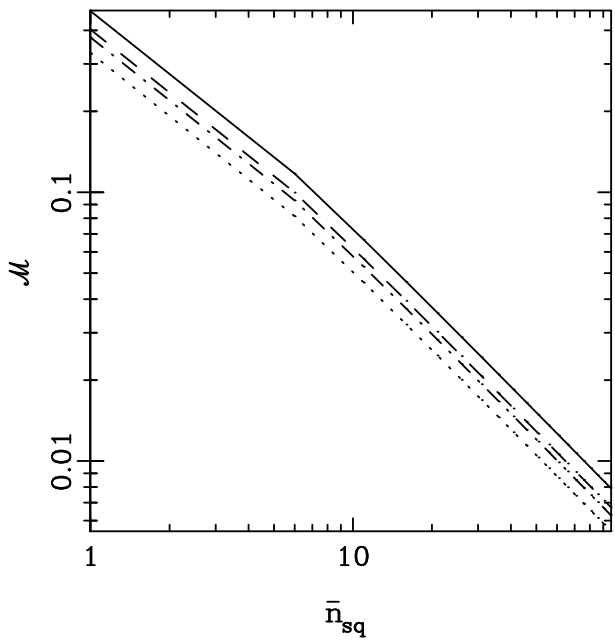,width=6cm}\end{center}
\caption{\sc The minimum detectable perturbation intensity ${\cal M}$ for 
the oscillator initially prepared in a squeezed vacuum and random phase 
perturbation. The behaviour of ${\cal M}$ is reported as a function 
of the mean excitation number $\bar{n}_{sq} = \sinh^2 r$ of the 
squeezed vacuum for different values of the false alarm probability 
$P_{01}$. The curves are clearly distinguishable in the region $\bar{n}_{sq} 
\sim 1$ where we have, from top to bottom, the behaviour for 
$P_{01}=0.00, 0.01, 0.02, 0.05$ respectively.} 
\label{f:minsq} 
\end{figure} 
\par 
In the case of random (unknown) phase we have not been 
able to solve analytically the perturbation equation. We solved it 
numerically. In Fig. \ref{f:minsq} we report the behaviour of $\cal M$ as a 
function of $\bar{n}_{sq}$ for various values of the false alarm probability. 
For $\bar{n}_{sq} \geq 10$ the numerical results are very well interpolated 
by the formula 
\begin{equation} 
{\cal M} \stackrel{\bar{n}_{sq} \geq 10}{\simeq} 
\log\left( \frac{2}{1+\sqrt{P_{01}(1-P_{01})}}\right) 
\frac{1}{\bar{n}_{sq}} 
\label{minsq3}\;. 
\end{equation}
\subsection{Number States} 
\label{sb:ns} 
The overlap of a number state with its displaced version is a real number, 
thus the overlap strength $\kappa$ is just the square of the overlap 
${\cal O}[n,z]$. We have 
\begin{equation} 
\kappa = \exp\{-|z|^2\} L_n^2 (|z|^2) 
\label{ovnum}\;, 
\end{equation} 
where $L_n (x)$ denotes a Laguerre polynomials \cite{grd}. 
Number states are a phase insensitive kind of states. Therefore, preparing the 
harmonic oscillator in such a way is equivalent to a phase averaging by default.\par
\begin{figure}[htb]  
\psfig{file=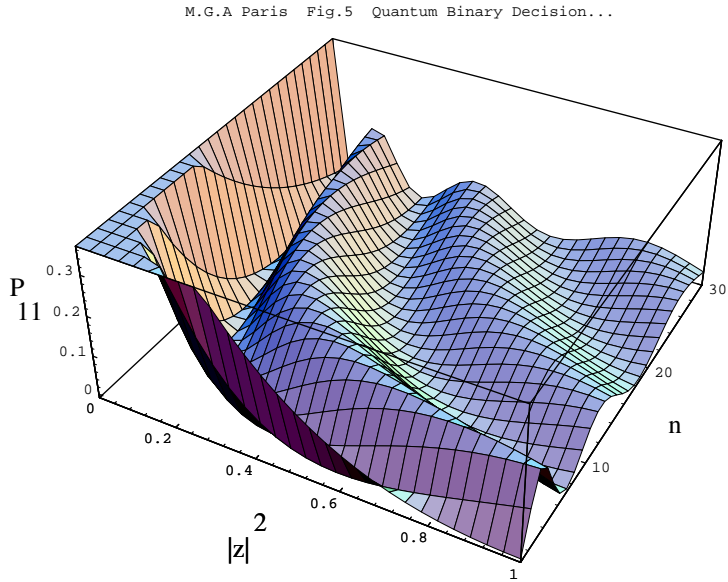,width=12cm}
\caption{\sc Three dimensional plot of the overlap strength $\kappa$ for 
the oscillator 
initially prepared in a number state. The overlap strength is reported as 
a function of the number $n$ and the perturbation intensity $|z|^2$.} 
\label{f:knum} 
\end{figure} 
\par 
In Fig. \ref{f:knum} we report the overlap 
strength as a function of the perturbation intensity and the excitation 
number of harmonic oscillator. \par 
The minimum perturbation equation (\ref{mincoh1}) reads as follows 
\begin{equation} 
\frac{1}{2}= \exp\{-{\cal M}\} L_n^2 ({\cal M}) 
\;. 
\end{equation} 
It could be numerically solved. Minimum detectable intensity scales 
as 
\begin{equation} 
{\cal M} \stackrel{n \gg 1}{\simeq} \frac{A}{n} 
\label{minnum}\;, 
\end{equation} 
with the proportionality constant depending on the value of the false alarm 
probability. Roughly we have 
\begin{equation} 
A \approx 0.3 - \frac{3}{2} P_{01} 
\label{minnum2}\;. 
\end{equation}
\subsection{Superposition of Coherent States} 
\label{sb:sch} 
We end this section by dealing with superposition states. We consider 
an analytically solvable case which is provided by superpositions of 
coherent states. Let us introduce the two set of states expressed by 
\begin{equation} 
|\psi_{\pm}\rangle = \frac{1}{2\sqrt{1\pm\exp\{-2|\alpha |^2\}}} \left( 
|\alpha\rangle \pm | - \alpha\rangle\right) 
\label{schdef}\;, 
\end{equation} 
where $| \alpha\rangle$ denotes a coherent state. These states are known also 
as even and odd Schr\"{o}edinger cats \cite{cat} as they are superposition 
states containing only even and odd number components respectively. 
The evaluation of the overlap can be carried out by means of the 
operatorial relations 
\begin{eqnarray} 
\hat D(-\alpha )\hat D(z) \hat D(\alpha )&=&\exp\{\bar{z}\alpha 
-\bar{\alpha }z\} \hat D(z) \nonumber\\ 
\hat D(\alpha )\hat D(z)\hat D(\alpha )&=&\hat D(z+2\alpha ) 
\label{oprel}\;. 
\end{eqnarray} 
We consider for simplicity $\alpha$ as a real number, thus we obtain 
\begin{eqnarray} 
{\cal O}[\alpha ,\pm ,z]&=& \frac{\exp\{ -\frac{1}{2}|z|^2\}} 
{1\pm\exp\{-2\alpha^2\}} 
\Big[\cos\left(2\alpha |z|\sin\varphi\right) 
\nonumber \\ 
&& \pm e^{-2\alpha^2} \cosh\left(2\alpha |z|\cos\varphi\right) 
\Big] 
\label{ovsch}\;, 
\end{eqnarray} 
being $\varphi$ the perturbation phase. 
After some calculations we arrive at the overlap strength for the fixed 
value $\varphi= 0,\pi /2$ and for the phase averaged case. We have 
\begin{eqnarray} 
\kappa_{0}^{\pm} &=& \frac{\exp\{-|z|^2\}}{(1\pm 2\exp\{-2\alpha^2\})^2} 
\times \nonumber\\ 
&\times& \Big[ 1\pm \exp\{-2\alpha^2\} \cosh\left( 2\alpha |z|\right)\Big]^2 
\label{ksch1}\;, 
\end{eqnarray} 
\begin{eqnarray} 
\kappa_{\pi/2}^{\pm} &=& \frac{\exp\{-|z|^2 
-4\alpha^2\}}{(1\pm2\exp\{-2\alpha^2\})^2} 
\times\nonumber\\ 
&\times& \Big[ 1\pm \exp\{2\alpha^2\} \cos\left( 2\alpha |z|\right)\Big]^2 
\label{ksch2}\;, 
\end{eqnarray} 
\begin{eqnarray} 
\bar{\kappa}^{\pm} &=& \frac{\exp\{-|z|^2\}}{(1\pm 2\exp\{-2\alpha^2\})^2} 
\times\nonumber\\ 
&\times& \Big[ J_0\left( 2\alpha |z|\right)  \pm \exp\{-2\alpha^2\} 
I_0\left( 2\alpha |z|\right)\Big]^2 
\label{ksch3}\;, 
\end{eqnarray} 
being $J_0 (x)$ and $I_0 (x)$ the Bessel function and the modified Bessel 
function of the first kind \cite{grd}. 
Notice that the mean 
excitation numbers for the superposition states are given by 
\begin{equation} 
\bar{n}_{\pm} = |\alpha |^2 
\frac{1\mp e^{-2|\alpha |^2}}{1\pm e^{-2|\alpha |^2}} 
\label{schnum}\;. 
\end{equation} 
The strong effect of phase matching is again apparent. The minimum 
perturbation equation can be solved in the asymptotic region of large 
excitation numbers, leading to the energy scaling 
\begin{equation} 
{\cal M}_0^{\pm} \stackrel{\bar{n}_{\pm}\gg 1}{\propto} 
\bar{n}_{\pm} 
\label{minsch1}\;, 
\end{equation} 
\begin{equation} 
{\cal M}_{\pi/2}^{\pm} \stackrel{\bar{n}_{\pm}\gg 1}{\propto} 
\frac{1}{2\bar{n}_{\pm}} 
\label{minsch2}\;, 
\end{equation} 
\begin{equation} 
{\cal M}_{rnd}^{\pm} \stackrel{\bar{n}_{\pm}\gg 1}{\propto} 
\frac{1}{\bar{n}_{\pm}} 
\label{minsch3}\;. 
\end{equation} 
In the case of random phase perturbation the minimum detectable 
intensity shows, at least asymptotically, an inverse scaling 
relative to the mean excitation number of the initial state. 
The same behaviour we have observed for squeezed vacuum 
and number state initial preparation, and this seems to indicate a 
general bound for detectability of perturbations. 
Actually, the minimum detectable intensity ${\cal M}$ becomes 
almost independent on the initial preparation in the limit 
of high excitation numbers. 
\section{Conclusions} 
\label{s:ou} 
Quantum detection and estimation theory has been applied in binary 
hypothesis testing, regarding possible perturbations on harmonic 
behaviour of a physical system. 
The action of an external driving signal is described by a displacement 
operator, thus excluding the possibility that the perturbed 
state of the oscillator could be orthogonal to that has been 
initially prepared. \par 
The detection probability 
has been evaluated, for different initial preparation of the oscillator, 
as a function of the perturbation intensity and the initial preparation 
energy. Minimum detectable perturbation intensities have been also evaluated, 
which represent the ultimate quantum limit in detecting a perturbation 
for fixed initial preparation of the harmonic oscillator. 
The lower bounds on detectable perturbation show a strong dependence 
on the phase matching. In-phase perturbations could be effectively 
detected only for weakly excited oscillators, as minimum detectable intensity 
linearly increases with initial energy. On the contrary, out-of-phase 
perturbations are easily detected also for high excitations. In the realistic 
case of random phase perturbation, the minimum detectable perturbation 
intensity seems to become independent on the initial preparation, at least 
in the asymptotic regime of large initial energy. This suggests that the 
inverse scaling ${\cal M}\propto \bar{n}^{-1}$ relative to the initial 
mean energy could be a general bound.\par 
An ultimate quantum limit for detection of perturbations 
could be defined independently from initial 
preparation, upon a further optimization over all the possible quantum states 
of the oscillator. 
\section*{Acknowledgments}
I would thank Valentina De Renzi for discussions and encouragments.
This work has been partially supported by "Francesco Somaini" Foundation.

\end{document}